\documentclass[prd,nofootinbib,floats,superscriptaddress,eqsecnum,tightenlines,preprintnumbers,11pt]{revtex4}

\usepackage{hyperref}
\usepackage{graphicx}
\usepackage{amsmath,amssymb,amsfonts,amsthm,latexsym}
\usepackage{marginnote}
\usepackage{color}

\def\be{\begin{equation}}
\def\ee{\end{equation}}
\def\beq{\begin{equation}}
\def\eeq{\end{equation}}
\newcommand{\bea}{\begin{eqnarray}}
\newcommand{\eea}{\end{eqnarray}}
\def\bi{\begin{itemize}}
\def\ei{\end{itemize}}
\def\ba{\begin{array}}
\def\ea{\end{array}}
\def\bfig{\begin{figure}}
\def\efig{\end{figure}}

\newcommand{\As}{A_*}

\newcommand{\aq}{\alpha}
\newcommand{\h}{\gamma}

\newcommand{\acc}{{\bf a}} 
\newcommand{\Y}{Y} 

\newcommand{\td}{{\rm tUd}}

\begin{document}

\preprint{YITP-18-17, IPMU17-0131}

\title{``Shadowy"  modes in
Higher-Order Scalar-Tensor theories}

\author{Antonio  De Felice}
\affiliation{Center for Gravitational Physics, Yukawa Institute for Theoretical Physics, Kyoto University, 606-8502, Kyoto, Japan}
\author{David Langlois}
\affiliation{Laboratoire Astroparticule et Cosmologie, CNRS, Universit\'e Paris Diderot Paris 7, 75013 Paris, France}
\author{Shinji Mukohyama}
\affiliation{Center for Gravitational Physics, Yukawa Institute for Theoretical Physics, Kyoto University, 606-8502, Kyoto, Japan}
\affiliation{Kavli  Institute  for  the  Physics  and  Mathematics  of  the  Universe  (WPI), \\
The  University  of  Tokyo  Institutes  for  Advanced  Study, \\
The  University  of  Tokyo,  Kashiwa,  Chiba  277-8583,  Japan}
\affiliation{Laboratoire de Math\'ematiques et Physique Th\'eorique (UMR CNRS 7350), Universit\'e Fran\c cois Rabelais, Parc de Grandmont, 37200 Tours, France}

\author{Karim Noui}
\affiliation{Laboratoire de Math\'ematiques et Physique Th\'eorique (UMR CNRS 7350), Universit\'e Fran\c cois Rabelais, Parc de Grandmont, 37200 Tours, France}
\affiliation{Laboratoire Astroparticule et Cosmologie, CNRS, Universit\'e Paris Diderot Paris 7, 75013 Paris, France}
\author{Anzhong Wang}
\affiliation{Institute  for  Advanced  Physics \& Mathematics,
Zhejiang  University  of  Technology,  Hangzhou  310032,  China}
\affiliation{GCAP-CASPER, Physics Department, Baylor University, Waco, TX 76798-7316, USA}

\date{\today}

\begin{abstract}
 We consider Higher-Order Scalar-Tensor theories which appear degenerate when restricted to the unitary gauge but 
 are not degenerate in an arbitrary gauge.
 We dub them U-degenerate theories. 
 We provide a full classification of theories that are either DHOST or U-degenerate and that are quadratic in second derivatives of the scalar field, and discuss its extension to cubic and higher order theories. Working with a simple example of U-degenerate theory, 
we find that, for configurations in which  the scalar field gradient is time-like, the apparent extra mode in such a theory can be understood as a generalized instantaneous, or ``shadowy" mode, which does not propagate. Appropriate boundary conditions, required by the elliptic nature of part of the equations of motion, lead to the elimination of the apparent instability associated with this extra mode. 

\end{abstract}

\maketitle

\section{Introduction}
Scalar-tensor theories have always played a prominent role in providing alternative theories of gravity.  During the last few years, special attention has been devoted to scalar-tensor theories whose Lagrangian contains second-order derivatives of a scalar field. An important requirement for such theories is the absence of any Ostrogradski ghost, i.e.\ an extra  degree of freedom generically associated with higher time derivatives. 

The absence of such a problematic extra mode is automatically guaranteed in Degenerate Higher-Order Scalar-Tensor (DHOST) theories introduced in \cite{Langlois:2015cwa,Langlois:2015skt},  for which the degeneracy of the Lagrangian 
leads to constraints that eliminate this potential extra scalar degree of freedom, even if the associated Euler-Lagrange equations are higher-order. DHOST theories were explicitly constructed up to quadratic order in \cite{Langlois:2015cwa} (see also \cite{Crisostomi:2016czh,deRham:2016wji,Achour:2016rkg} for further details) and their full classification up to cubic order (in second derivatives) was completed in \cite{BenAchour:2016fzp}. DHOST theories extend the  class of Horndeski theories~\cite{Horndeski:1974wa}  and  the (larger) class of Beyond Horndeski theories~\cite{Gleyzes:2014dya,Gleyzes:2014qga} (another special subclass of DHOST theories was found in \cite{Zumalacarregui:2013pma}, via disformal transformations of the Einstein-Hilbert action).

In order to study Higher-Order Scalar-Tensor (HOST) theories, it is often convenient to resort to the so-called unitary gauge, where the coordinates are chosen such that the scalar field is spatially uniform, i.e.\ with only a time dependence. In other words, the constant time hypersurfaces coincide with the constant scalar field hypersurfaces. This gauge choice is of course restricted to configurations where the gradient of the scalar field is time-like but this is a natural assumption in the cosmological context. In particular, the unitary gauge is a key ingredient of the effective description of modification of gravity, dark energy and inflation (see e.g. 
 \cite{ArkaniHamed:2003uy,ArkaniHamed:2003uz,Creminelli:2006xe,Cheung:2007st,Gubitosi:2012hu,Gleyzes:2013ooa,Gleyzes:2014rba} and especially \cite{Langlois:2017mxy}  devoted to DHOST theories). 

For Beyond Horndeski theories, the counting of the number of degrees of freedom was initially carried out via a Hamiltonian formulation in the unitary gauge~\cite{Gleyzes:2014dya,Gleyzes:2014qga,Lin:2014jga}. 
Potential limitations of the unitary gauge were later pointed out in \cite{Deffayet:2015qwa}, where a Hamiltonian analysis valid in an arbitrary gauge was also presented for a particular Beyond Horndeski theory (which in fact is related to a Horndeski theory by a disformal transformation, according to the correspondence shown earlier in \cite{Gleyzes:2014qga}). A Hamiltonian analysis in an arbitrary gauge, using explicitly the degeneracy of the Lagrangian, for all quadratic HOST (including DHOST) theories was subsequently given in \cite{Langlois:2015skt}. 

A manifest pitfall of the unitary gauge is that there exist HOST theories which seem to be degenerate when written in the unitary gauge but are not
degenerate in their fully covariant version and therefore are not DHOST theories. 
We will denote these theories U-degenerate. The purpose of the present work is to study this very special class of theories and better understand the number and role of the scalar degrees of freedom, from the point of view of the unitary gauge or from that of an arbitrary gauge.

In this work, we first 
 present a systematic and simple way to classify all HOST theories that are either DHOST or U-degenerate. For quadratic theories (in second derivatives of the scalar field), we find that their Lagrangian $L$ can be written as the sum of a totally U-degenerate Lagrangian,  by which we mean   a Lagrangian whose kinetic terms (for the scalar and tensor modes) vanish in the unitary gauge,  and another term that does not involve the metric curvature and can be written in a simple way that makes the degeneracy in the unitary gauge manifest. Both terms of the Lagrangian  correspond to DHOST Lagrangians separately, but their sum is not a DHOST Lagrangian. We then  generalize this result  to Lagrangians that involve arbitrary powers of second derivatives 
 $\phi_{\mu\nu}\equiv \nabla_\nu\!\nabla_\mu\phi$, starting with cubic theories. This provides  a  simple and systematic parametrization of theories that  are either DHOST or U-degenerate.

Interestingly, U-degenerate HOST theories include as particular examples the khronometric theories discussed in \cite{Blas:2010hb,Blas:2011ni}. For these theories, the extra  mode that appears in the covariant formulation has been called ``instantaneous mode.'' 
 In the more general context that we consider here, the structure of the extra mode that appears is often more intricate than in the case of ``instantaneous'' modes. We will call this mode a ``generalized instantaneous mode'',  or also ``shadowy'' mode for a shorter denomination. 

The notion of generalized instantaneous or shadowy mode can easily be understood by considering  the following example of a non-dynamical Lagrangian in Minkowski spacetime, 
\begin{equation}
L[\psi]=\frac12 \psi \Delta \psi \, ,
\end{equation}
 where $\Delta$ is the Euclidean Laplacian operator. This Lagrangian leads to  the Laplace equation
$\Delta \psi=0$.  
In a different set of coordinates $(t',x',y',z')$,  with $t' = t+vx$ $(v\ne 0)$ and the same spatial coordinates, the Lagrangian for $\psi$ becomes  
\begin{equation}
 L[\psi] = -\frac{1}{2}\left[(v\partial_{t'}\psi+\partial_{x'}\psi)^2+(\partial_{y'}\psi)^2+(\partial_{z'}\psi)^2\right] \quad \ni \ -\frac{v^2}{2}(\partial_{t'}\psi)^2\,, \label{eqn:Lchi_tilted}
\end{equation}
which contains a kinetic term for $\psi$ (with a negative sign).
In this new frame, the action seems to contain a dynamical degree of freedom, which corresponds to a shadowy mode.

In order to better understand  the ``shadowy'' mode that arises in U-degenerate HOST theories, we study in detail a simple toy-model. It is a higher-derivative scalar theory, inspired from U-degenerate HOST theories, which we study in a flat two-dimensional spacetime for simplicity. We consider some background solution and then make a linear perturbation analysis around this background solution in two different coordinate systems. In the first one, the background solution is only time-dependent, corresponding to the choice of the unitary gauge
for the background.
In the second one, the background solution is both time and space dependent, 
but the gradient of the background scalar field  is still assumed to be time-like. 
We then identify, in both approaches, the  degrees of freedom of the system and study the correspondence between these two calculations.

We find that the extra mode (which appears when the background is time and space dependent) can be understood as a shadowy mode, which does not really propagate. Appropriate boundary conditions, required by the elliptic nature of part of the equations of motion, lead to  the elimination of the apparent instability associated with this extra mode. 
Hence, our analysis in this simple toy-model reconciles the two seemingly contradictory points of view based on the unitary gauge and a non-unitary gauge.
This toy model also illustrates that the unitary gauge (which can be used for configurations where the gradient of the scalar field is  time-like) 
 constitutes a  convenient gauge choice, where the partially elliptic character of the equations of motion is more transparent and  where it is thus easier to fix appropriate boundary conditions. 

\medskip

The paper is organized  as follows. In section \ref{Udegsection},  we present the classification
of HOST theories that are U-degenerate, first focussing on quadratic theories then extending our classification to higher order.  In section \ref{illustrative example},  we study in detail a simple but illustrative example of U-degenerate theory in a two-dimensional Minkowski spacetime and analyse the number and nature of degrees of freedom, depending on the gauge chosen to describe the background solution.  We conclude 
with a brief summary and a discussion.
Some technical details are also given  in the Appendices.

\section{U-degenerate HOST theories}
\label{Udegsection}
The goal of this section is to present a classification of U-degenerate HOST theories, i.e.\ Higher-Order Scalar-Tensor (HOST) theories that are degenerate only in the unitary gauge. For this purpose, 
we actually provide  a classification of theories that are either DHOST or U-degenerate, i.e. those that are degenerate at least in the unitary gauge. 
After a short review of DHOST theories, which enables us to introduce some useful notations,  we classify HOST Lagrangians that are either DHOST or U-degenerate and that are quadratic in second derivatives $\phi_{\mu\nu}$. 
We then extend our classification to cubic theories and beyond in the last two subsections.

\subsection{DHOST theories}
\label{dhostrebiew}
We start with HOST theories whose Lagrangian is (at most) quadratic in the second derivatives of the scalar field.
The action of these theories takes the form
\bea\label{general action}
S[g_{\mu\nu},\phi] \; = \; \int d^4x \sqrt{-g} \, \left[ f_2(\phi,X) \, R \; + \; L^{(2)}_\phi +f_0(\phi,X)+f_1(\phi,X)\Box \phi  \right] \, ,
\eea
where $R$ is the Ricci scalar, $f_A(\phi,X)$ are arbitrary functions of $\phi$ and $X\equiv \phi_\mu \phi^\mu$ with
$\phi_\mu \equiv \nabla_\mu \phi$. 
The term $L^{(2)}_\phi$  denotes the most general minimal coupling Lagrangian quadratic in $\phi_{\mu\nu} \equiv \nabla_\mu \phi_\nu$ and is given by
\bea\label{Lphi}
L^{(2)}_\phi \; \equiv \; \sum_A \alpha_A(\phi,X) \, L^{(2)}_A \, ,
\eea
 where $\alpha_A(\phi,X)$ are functions of $\phi$ and $X$, and the elementary quadratic Lagrangians $L^{(2)}_A$ are 
\bea\label{elem}
&&L^{(2)}_1 \; = \; \phi_{\mu\nu} \phi^{\mu\nu} \,,  \quad
L^{(2)}_2 \; = \; (\Box \phi)^2 \, , \quad
L^{(2)}_3 \; = \; (\phi^\mu\phi^\nu \phi_{\mu\nu}) \Box \phi \, , \nonumber \\
&&L^{(2)}_4 \; = \; (\phi_{\mu\nu}\phi^\nu \phi^{\mu\sigma}\phi_\sigma) \, , \quad
L^{(2)}_5 \; = \; (\phi^\mu\phi^\nu \phi_{\mu\nu})^2 \,.
\eea

These theories  can be extended to include  cubic terms,  by adding to the action \eqref{general action} the terms
\bea
 \int d^4x \, \sqrt{-g} \, \left( f_3(\phi,X) \phi_{\mu\nu} \, G^{\mu\nu} \, + \, \sum_{A=1}^{10} b_A(\phi,X) L^{(3)}_A \right) \, ,
\eea
where the ten elementary cubic Lagrangians $L_A^{(3)}$ are \cite{BenAchour:2016fzp}
\be
\label{CubicL}
\begin{split}
& L^{(3)}_1=  (\Box \phi)^3  \,, \quad
L^{(3)}_2 =  (\Box \phi)\, \phi_{\mu \nu} \phi^{\mu \nu} \,, \quad
L^{(3)}_3= \phi_{\mu \nu}\phi^{\nu \rho} \phi^{\mu}_{\rho} \,,   \\
& L^{(3)}_4= \left(\Box \phi\right)^2 \phi_{\mu} \phi^{\mu \nu} \phi_{\nu} \,, \quad
L^{(3)}_5 =  \Box \phi\, \phi_{\mu}  \phi^{\mu \nu} \phi_{\nu \rho} \phi^{\rho} \,, \quad
L^{(3)}_6 = \phi_{\mu \nu} \phi^{\mu \nu} \phi_{\rho} \phi^{\rho \sigma} \phi_{\sigma} \,,   \\
& L^{(3)}_7 = \phi_{\mu} \phi^{\mu \nu} \phi_{\nu \rho} \phi^{\rho \sigma} \phi_{\sigma} \,, \quad
L^{(3)}_8 = \phi_{\mu}  \phi^{\mu \nu} \phi_{\nu \rho} \phi^{\rho}\, \phi_{\sigma} \phi^{\sigma \lambda} \phi_{\lambda} \,,   \\
& L^{(3)}_9 = \Box \phi \left(\phi_{\mu} \phi^{\mu \nu} \phi_{\nu}\right)^2  \,, \quad
L^{(3)}_{10} = \left(\phi_{\mu} \phi^{\mu \nu} \phi_{\nu}\right)^3 \,.
\end{split}
\ee

In general, these theories propagate two scalar modes in addition to the usual two tensorial modes, one of the two scalar modes being an Ostrogradsky mode. However, when the Lagrangian is degenerate (i.e.\ it admits at least one primary constraint in addition to the usual constraints associated with the diff-invariance), the theory propagates at most three degrees of freedom: the extra constraints enable us to eliminate some degrees of freedom. The classification of degenerate theories up to cubic order has been completed in \cite{BenAchour:2016fzp}. 

\medskip

\subsection{Classification of U-degenerate quadratic Lagrangians}
\label{quad}
In order to classify all Lagrangians that are either DHOST or U-degenerate, it is useful to start from  the ADM decomposition of \eqref{general action} in the unitary gauge, ignoring the $f_0$ and $f_1$ terms which do not play any role in the degeneracy. 
We thus  write the four-dimensional metric in the form
\be
ds^2=-N^2 dt^2+\gamma_{ij}\left(dx^i+N^i dt\right)\left(dx^j+N^j dt\right)\,,
\ee
where $N$ and $N^i$ are the lapse and shift, respectively, and $\gamma_{ij}$ is the 3-dimensional metric on constant $t$ spatial hypersurfaces. In the following, a dot will denote a partial derivative with respect to the time coordinate $t$.  

As shown in \cite{Langlois:2015skt}, the kinetic part of the (3+1) decomposition of the action (\ref{general action}) can be written in the form
\bea\label{3+1decomp}
S_{\rm kin}  & = & \int  dt \, d^3x \, N \sqrt{\h} \left({\cal A} \, \dot\As^2 +
2 {\cal B}^{ij} \dot\As K_{ij} + {\cal K}^{ijkl} K_{ij} K_{kl}\right)\,,
 \eea 
where  $K_{ij}$ is the extrinsic curvature tensor and 
 \beq
 \As\equiv \frac1N\left(\dot\phi-N^i\partial_i \phi\right)\,.
 \eeq
In the unitary gauge (where $\partial_i\phi=0$), the coefficients  that appear in \eqref{3+1decomp} reduce  to
\bea
\label{kin_coeff1}
{\cal A}_{\rm U} &=&\aq_1 +\aq_2 +(\aq_3 + \aq_4) X_U + \aq_5 X_U^2  \, , \\
{\cal B}^{ij}_{\rm U} &=&4 f_{2X} +2 \aq_2 +\aq_3 X_U \,, \\
 {\cal K}^{ij,kl}_{\rm U}&=&\left(f_2 -\aq_1 X_U\right) \h^{i(k}\h^{l)j} -\left( f_2 -\aq_2 X_U\right) \h^{ij} \h^{kl}\, \,,
 \label{kin_coeff3}
\eea
with $X_{\rm U}\equiv-\As^2=-\dot\phi^2/N^2$, corresponding to the expression of $X$ in the unitary gauge.
The full expressions of these coefficients in an arbitrary gauge can also be found in  \cite{Langlois:2015skt}, but we will not need them here. 

Let us first identify the Lagrangians that are   non-dynamical when restricted to the unitary gauge, i.e.\ for which all of the above coefficients vanish. 
As one can immediately see, this imposes four conditions on the six functions $f_2$ and $\aq_A$. 
Thus, the family of Lagrangians which are non-dynamical, i.e. totally U-degenerate,  in the unitary gauge can be expressed in terms of only two free functions (4 conditions for 6 initial free functions), for instance $f_2$ and $\aq_5$, while  the other four are determined   by the relations
\bea
\label{degcond}
\aq_1=-\aq_2 =\frac{f_2}{X} \, , \quad
\aq_3= \frac{2}{X} \left( \frac{f_2}{X} - 2 f_{2X}\right)\, , \quad \aq_4 =  \frac{2}{X} \left(2 f_{2X} - \frac{f_2}{X} \right) -X \aq_5 \, .
\eea
This means that the   quadratic Lagrangians that are totally U-degenerate in the unitary gauge can explicitly be written in the form
\bea\label{LU0}
L_{\td}[f_2, \aq_5] & \equiv & f_2 \, R + \frac{f_2}{X} \left(L_1^{(2)}-L_2^{(2)}\right) \nonumber \\
&+& \frac{2}{X^2} \left(f_2-2Xf_{2X}\right) \left(L_3^{(2)}-L_4^{(2)}\right) - \aq_5 \left(X L_4^{(2)} -  L_5^{(2)}\right) \, , ~~~~~
\eea
where $f_2$ and $\aq_5$ are free functions.

In order to classify all quadratic HOST theories that are U-degenerate, it is convenient to decompose any Lagrangian into a totally U-degenerate part \eqref{LU0}, which includes the Ricci scalar term, and another part which depends only on the five elementary Lagrangians of  (\ref{elem}). The total Lagrangian thus reads
\bea
\label{L0Lphi}
L \; = \; L_\td[f_2,0] + \tilde L_\phi \,,
\eea
where $\tilde L_\phi$ is of the form (\ref{Lphi}).
As already mentioned, the  $f_0$ and $f_1$ terms are not taken into account here because they do not modify the degeneracy properties of the total Lagrangian.

Since the kinetic part of $L_\td[f_2,0]$ vanishes in the unitary gauge, it is easy to see that any Lagrangian $L$ is U-degenerate  if and only if the Lagrangian $\tilde L_\phi$ is also U-degenerate. 
Moreover, degeneracy of $\tilde L_\phi$ means that the kinetic part of the Lagrangian, in the unitary gauge, 
can be written in  the form
\bea\label{factor}
\tilde{L}_{\phi,{\rm kin}} \; = \;\hat{\cal K}^{ij,kl}_{\rm U} \left( K_{ij} + \sigma  \gamma_{ij} \dot\As\right)
 \left( K_{kl} + \sigma  \gamma_{kl} \dot\As \right) \,,
 \eea
 where 
 \beq
 \hat{\cal K}^{ij,kl}_{\rm U}=  -X_{\rm U}\, \left(\aq_1 \,  \h^{i(k}\h^{l)j} -\aq_2 \,  \h^{ij} \h^{kl}\right) \,,
 \eeq
which corresponds to (\ref{kin_coeff3}) with $f_2=0$,  since $\tilde L_\phi$ does not contain any curvature term by construction. 
 By expanding (\ref{factor}) and comparing with (\ref{3+1decomp}),  one finds (by eliminating $\sigma$) that the U-degenerate form (\ref{factor}) is
 possible if and only if the functions $\alpha_A$ satisfy the relation
\bea\label{Udeg}
4(\alpha_1+3\alpha_2) \left(\alpha_1 + \alpha_2 + X(\alpha_3+\alpha_4)  + X^2\alpha_5 \right) \; = \; {3(2\alpha_2 + X \alpha_3)^2} \,.
\eea
Not surprisingly, it coincides  with the degeneracy condition in the unitary gauge, already derived in \cite{Langlois:2015cwa}. 
Note that, by definition, U-degenerate theories satisfy the condition (\ref{Udeg}) but not all three degenerate conditions obtained in \cite{Langlois:2015cwa}. 

\medskip

The expression of the Lagrangian written in the unitary gauge can easily  be ``covariantized" by using the Stueckelberg trick (see the Appendix for the correspondence). One thus obtains, instead of the parametrization in terms of the functions $\aq_A$, a parametrization of all U-degenerate theories in terms of the five functions $f_2$, $\kappa_1$, $\kappa_2$, $\sigma$ and $\alpha$,  which depend on $X$ and $\phi$, with  a Lagrangian of the form 
\bea\label{LUdegcov}
L  =  L_\td[f_2,\alpha]  +\hat{\cal K}^{\mu\nu,\rho\sigma} \left( \phi_{\mu\nu} + \sigma \, \Y\,  g_{\mu\nu}\right) 
\left( \phi_{\rho\sigma} + \sigma \, \Y\, g_{\rho\sigma}\right) \, ,
\eea
where
\bea
\hat{\cal K}^{\mu\nu,\rho\sigma}  \; \equiv \; \kappa_1 h^{\mu(\rho} h^{\nu)\sigma} + \kappa_2 h^{\mu\nu} h^{\rho\sigma} \,, \qquad 
h_{\mu\nu} \equiv g_{\mu\nu}- \frac{1}{X} \, \phi_\mu \phi_\nu \,,\qquad \Y\equiv \phi^\alpha \phi_{\alpha\beta} \phi^\beta\,.
\eea
This is of course compatible with the parametrization \eqref{L0Lphi} in terms of six functions constrained
by the single relation \eqref{Udeg}. To make the relationship between these two parametrizations explicit, let us expand
\eqref{LUdegcov} in terms of the elementary Lagrangians. One obtains
\bea
\label{L_Ud_quad}
L & = & f_2 \, R + \left( \kappa_1+\frac{f_2}{X}  \right) L^{(2)}_1
+ \left( \kappa_2 - \frac{f_2}{X}\right)L^{(2)}_2 \nonumber \\
&+ &\left( 2 \frac{f_2}{X^2} - 4 \frac{f_{2X}}{X} + 2\sigma \kappa_1 + 2 \left[3\sigma - \frac{1}{X}\right]\kappa_2\right)L^{(2)}_3 \nonumber \\
&+ & \left( \alpha + 2\frac{f_{2X}}{X} - \frac{2f_2}{X^2} - \frac{2}{X} \kappa_1 \right)L^{(2)}_4 \nonumber \\
&+ &\left( -\frac{\alpha}{X} + \frac{2f_{2X}}{X^2} + \kappa_1\left[\frac{1}{X^2} + 3\sigma^2 - \frac{2\sigma}{X}\right] + \kappa_2 \left[3\sigma - \frac{1}{X}\right]^2\right)L^{(2)}_5 \, ,
\eea
and it is straightforward to check that this Lagrangian indeed satisfies the condition \eqref{Udeg}. 
The above Lagrangian includes both U-degenerate theories and DHOST theories, since the latter also satisfy the condition \eqref{Udeg}.

\subsection{Beyond quadratic order}
\label{cubic}
In order to classify all theories that are either DHOST or U-degenerate up to third order in second derivatives of $\phi$, one can follow the same strategy as in the previous section and first identify the theories that are totally U-degenerate, i.e.\ nondynamical in the unitary gauge. By using the ADM decomposition of HOST theories, up to cubic order, given in  \cite{Langlois:2017mxy}, one finds that all the kinetic terms vanish in the unitary gauge when the 
 following eleven relations  are satisfied by the functions $\aq_A$, $b_A$ and $f_A$ :
\bea
&&b_1=b_2=b_3=0 \, , \quad
b_4=-b_6=-\frac{f_{3X} }{X} \, , \quad b_5+Xb_9 = -\frac{2f_{3X} }{X}  \, , \nonumber \\
&&  b_7+Xb_8+X^2 b_{10}=\frac{2f_{3X} }{X}  \, , \quad \aq_1=\frac{f_2}{X} + \frac{f_{3\phi}}{2} \, , \quad
\aq_2=-\frac{f_2}{X} + \frac{f_{3\phi}}{2} \, , \\
&& \aq_3 =\frac{2}{X} \left( \frac{f_2}{X} - \frac{f_{3\phi}}{X} - 2 f_{2X}\right) \, , \quad
\aq_4+X\aq_5=\frac{2}{X} \left( 2 f_{2X} - \frac{f_2}{X}\right) . \nonumber
\eea
Since the initial Lagrangian depends on $17$  functions, this implies that  totally U-degenerate theories  depend on  $6$ arbitrary functions that can be chosen to be $f_3,b_8,b_9, b_{10}$ for the cubic part, and $f_2,\aq_5$, as before  for the quadratic part, so that 
\beq
\label{Ltd}
L_\td[f_2, f_3, \aq_5, b_8,b_9, b_{10}]=L^{(2)}_\td[f_2, f_3, \aq_5]+L^{(3)}_\td[f_3, b_8,b_9, b_{10}]\,,
\eeq
where, on the right-hand side, we have separated the terms that can be expressed in terms of the scalar curvature and of the quadratic Lagrangians (\ref{elem}), and those written in terms of $G^{\mu\nu}\phi_{\mu\nu}$ and  (\ref{CubicL}). Note that the quadratic part depends on $f_3$ too, if $f_{3\phi}$ is nonzero. 

Similarly to the quadratic case discussed previously, all  U-degenerate Lagrangians up to cubic order can be written in the form 
\bea
L & = & L_{\td}[f_2, f_3, \aq_5, b_8,b_9, b_{10}]+\hat{\cal K}^{\mu\nu,\rho\sigma}  \left( \phi_{\mu\nu} + \sigma \Y g_{\mu\nu}\right) 
\left( \phi_{\rho\sigma} + \sigma \Y  g_{\rho\sigma}\right) \nonumber\\
&+& \hat{\cal K}_3^{\mu\nu,\rho\sigma,\alpha\beta}  \left( \phi_{\mu\nu} + \sigma \Y g_{\mu\nu}\right) 
\left( \phi_{\rho\sigma} + \sigma \Y g_{\rho\sigma}\right)
\left( \phi_{\alpha\beta} + \sigma \Y  g_{\alpha\beta}\right)\; ,  \label{cubiccovdeg}
\eea
where 
\bea
{\cal K}_3^{\mu\nu,\rho\sigma,\alpha\beta}  & \equiv & 
\omega_1 \, h^{\mu\nu} h^{\rho\sigma} h^{\alpha\beta} +  \omega_3 \, \left(  h^{\nu(\rho} h^{\sigma)(\alpha} h^{\beta)\mu} +  h^{\mu(\rho} h^{\sigma)(\alpha} h^{\beta)\nu}\right)\nonumber \\
&&+ {\omega_2} \, \left(h^{\mu(\rho} h^{\sigma)\nu} h^{\alpha\beta} + h^{\alpha(\rho} h^{\sigma)\beta} h^{\mu\nu} + h^{\alpha(\mu} h^{\nu)\beta} h^{\rho\sigma} \right)  \, ,
\eea
with $\sigma$ and $\omega_A$ arbitrary functions of $\phi$ and $X$.
One can show that the three parts of the Lagrangian  \eqref{cubiccovdeg} correspond separately to  DHOST theories. 
In fact, the last two terms of (\ref{cubiccovdeg}), which do not depend on $f_2$ and $f_3$, correspond to any 
DHOST Lagrangian satisfying $f_2=f_3=0$. They have been classified 
in \cite{BenAchour:2016fzp}, but can also be written in this very simple form, parametrized by $10$ arbitrary functions of $X$ and $\phi$, namely  $\aq_5, b_8,b_9, b_{10}$, $\kappa_1$, $\kappa_2$,  $\omega_1$, $\omega_2$, $\omega_3$ and $\sigma$.  One can  verify that these Lagrangians indeed satisfy
the degeneracy conditions presented in \cite{BenAchour:2016fzp}.

One can then generalize these results to parametrize U-degenerate theories  with a  Lagrangian
$L$ that contains arbitrary powers of $\phi_{\mu\nu}$. Following (\ref{cubiccovdeg}), one writes  $L$  as  $L=L_\td+L_\phi$
where $L_\td$ is given by (\ref{Ltd}) and contains all the curvature terms, while  $L_\phi$ is a  degenerate Lagrangian obtained by combining 
$X_{\mu\nu}\equiv \phi_{\mu\nu} + \sigma Y g_{\mu\nu}$ with the projector $h_{\mu\nu}$, 
\bea
L_\phi \; = \; {\cal K} \left[ \phi_{\mu\nu} + \sigma\,  Y g_{\mu\nu}, h^{\rho\sigma}\right] \,.
\eea
Formally, one can expand $\cal K$  as
\bea
{\cal K}(X_{\mu\nu}) \; = \; \sum_{A} {\cal K}_A^{\mu_1\nu_1,\mu_2\nu_2,\cdots,\mu_A \nu_A} \, X_{\mu_1\nu_1} \cdots X_{\mu_A\nu_A} \, ,
\eea
where ${\cal K}_A$ are tensors constructed from $h^{\mu\nu}$ only. 
Once again, let us stress that the general Lagrangians given above include both U-degenerate theories and DHOST theories.

\section{An illustrative example}
\label{illustrative example}
After having classified all U-degenerate theories in the previous section, we would  like to better understand  the number of degrees of freedom present in these theories, as well as their nature. In particular, since U-degenerate theories are degenerate in the unitary gauge but non-degenerate in another gauge, one would naively expect the presence of a single scalar degree of freedom in the unitary gauge but the appearance of an extra scalar degree of freedom when working in another gauge. We would like to understand how these two seemingly contradictory points of view can be reconciled.

For simplicity, we are going to restrict our analysis to a very simple model, directly inspired from the classification of the previous section but for which we ignore the tensor degrees of freedom to concentrate only on the scalar modes. Such a simple model is provided, for instance, by the totally U-degenerate Lagrangian $L_\td[0, \mu]$ defined in (\ref{LU0}), where we choose $\mu$ to be constant, restricted to a Minkowski spacetime. This Lagrangian is however too simple in the sense that it does not contain any propagating degree of freedom in the unitary gauge, where it is totally degenerate. For this reason, we add to this Lagrangian  a standard kinetic term, which guarantees the presence of a propagating degree of freedom in the unitary gauge. 

We  thus consider the following Lagrangian
\beq
\label{L_ex}
L=-\frac12 X-\mu \left(X L^{(2)}_4-L^{(2)}_5\right)\,.
\eeq 
Since we do not consider the metric fluctuations, the kinetic part of the Lagrangian   written in the ADM form (\ref{3+1decomp})  reduces, ignoring $f_0$ 
(in the present case, $-X/2$), to
\beq 
L_{\rm kin}= {\cal A} \, \dot\As^2\,,
\eeq
where
\beq
{\cal A} =\aq_1 +\aq_2 -(\aq_3 + \aq_4) \As^2 + \aq_5 \As^4= \mu \left(X\As^2+\As^4\right)\,.
\eeq
In the unitary gauge, $X=-\As^2$ and ${\cal A}$ vanishes. However, in an arbitrary gauge, we find 
\beq
{\cal A} =\mu \, \As^2\,  (\partial_i\phi)^2,
\eeq
which does not vanish in general. We would thus expect to find an extra mode in this case.

\subsection{Analysis in a unitary gauge background}
Assuming that the scalar field has a time-like spacetime gradient, we can work in the unitary gauge and consider the background field
\beq
\bar\phi=t\,.
\eeq
Considering the perturbed solution
\beq
\phi=t+\chi(t,x)\,,
\eeq
the Lagrangian quadratic in perturbations is given by 
\beq
\label{L_0}
{\cal L}_0=\frac12 \left(\dot\chi^2-\chi^{\prime 2}\right)+\mu \dot\chi^{\prime 2} \, ,
\eeq
and no second time derivative appears. This is to be expected since the background is in the unitary gauge.

The dispersion relation is 
\beq
\label{dispersion_0}
 \left(1+2 \mu k^2 \right)  
   \omega ^2
-k^2=0\,,
\eeq
which gives the two solutions 
\beq
\omega=\pm \frac{k}{\sqrt{2\mu k^2+1}}\,,
\eeq
corresponding to a single degree of freedom. Here, as a boundary condition, we have implicitly assumed that the field does not diverge at spatial infinity so that $k$ is real.

\subsection{Analysis in a non-unitary gauge background}

We now consider a background solution of the form
\beq
\label{phi_bck}
\bar\phi=t+\alpha x\,,
\eeq
which is also a solution of the equations of motion. If $\alpha\neq 0$, this background solution is not described in the unitary gauge since the scalar field has now an explicit spatial dependence.

We then consider the perturbed solution
\beq
\phi=t+\alpha x+\chi(t,x)\,.
\eeq
Substituting into (\ref{L_ex}), one can derive the Lagrangian quadratic in $\chi$, which reads
\begin{eqnarray}
\label{Lalpha}
{\cal L}_\alpha&=&\frac12 \left(\dot\chi^2-\chi^{\prime 2}\right)+\mu\left[\alpha^2\left(\ddot \chi^2+\chi^{\prime\prime 2}\right)-2\alpha(1+\alpha^2) (\ddot\chi\dot\chi'+ \chi''\dot\chi')
+(1+4\alpha^2+\alpha^4) \dot\chi^{\prime 2}
\right]\,. \quad 
\end{eqnarray}
One can immediately check that, when $\alpha=0$, one recovers the previous case (\ref{L_0}).

\begin{figure}
\begin{center}
\includegraphics[width=4in]{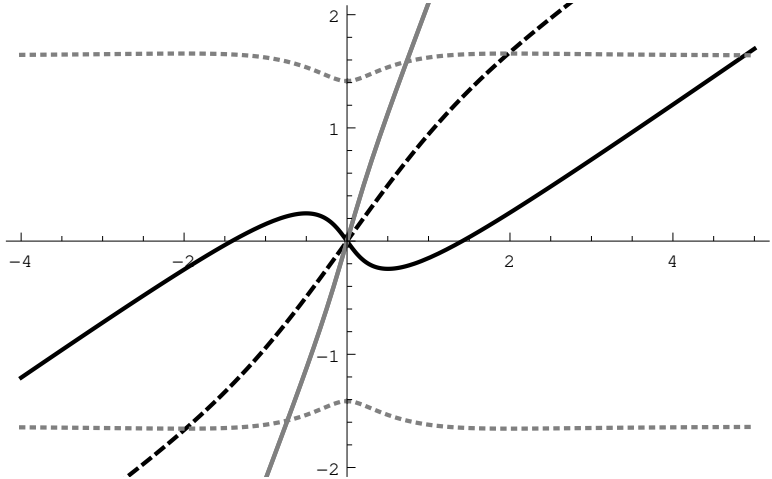}
\end{center}
\caption{Frequency $\omega$ as a function of $k$ for the four modes that appear in the non-unitary gauge: $\omega_1(k)$, $\omega_2(k)$ correspond respectively to the dashed and continuous black curves. The real and  imaginary parts of $\omega_3(k)$ and $\omega_4(k)$ are plotted respectively as continuous and dotted grey curves.}
\label{fig}
\end{figure}

By considering plane wave solutions of the equations of motion,  of the form $\chi\propto \exp(-i\omega t+i k x)$, one obtains the dispersion relation
\beq
\label{dispersion}
2 \alpha ^2 \mu \, \omega ^4
+4 \alpha 
   \left(\alpha ^2+1\right) k \mu \, 
   \omega ^3
  + \left(2
   \left(\alpha ^4+4 \alpha
   ^2+1\right) k^2 \mu
   +1\right)  
   \omega ^2
 +4 \alpha 
   \left(\alpha ^2+1\right) k^3 \mu\,
    \omega +2 \alpha
   ^2 k^4 \mu-k^2=0\,.
\eeq
In contrast with (\ref{dispersion_0}), this dispersion relation is polynomial in  $\omega$  up to fourth order, when $\alpha\neq 0$ (if $\alpha=0$, one recovers (\ref{dispersion_0}) obviously). This leads to four solutions for $\omega$: two of them are real and we will denote them $\omega_1$ and $\omega_2$. The other two are complex conjugate, i.e.\ of the form 
\beq
\omega_{3,4}=\omega_\pm= \omega_r\pm i\,  \omega_i\,.
\eeq
The four solutions of the dispersion relation (\ref{dispersion}), for a particular choice of $\alpha$ and $\mu$,  are plotted in Fig.\ \ref{fig}. 

A priori, the fact that the equation of motion is fourth order indicates that four initial conditions need to be specified to evolve the system. At some initial time, say $t=0$, one needs as initial data the four functions $\chi(0,x)$, $\dot\chi(0,x)$, $\ddot\chi(0,x)$, $\dddot\chi(0,x)$,  which can be assumed to be regular and to decay at spatial infinity (or even be nonzero only in a compact region of space). 

In Fourier space, the equation for $\chi$  yields 
an ordinary differential equation for each Fourier mode $\chi(t, k)$, which is fourth order in time derivatives. This equation admits four independent particular solutions of the form $\chi(t, k)\propto e^{i\omega t}$, corresponding to the four solutions  $\omega$ of the dispersion relation (\ref{dispersion}). As a consequence, the general solution can be written in the form
\beq
\label{chi_evolution}
\chi(t, k)=\sum_{A=1}^4 u_A(k)\ e^{i\,\omega_A(k)  t}\,,
\eeq
where the coefficients $u_A(k)$ are determined from the initial conditions $\chi(0,x)$, $\dot\chi(0,x)$, $\ddot\chi(0,x)$, $\dddot\chi(0,x)$, or equivalently $\chi^{(n)}(0,k)$ for $n=0,1,2,3$, by inverting the four relations 
\beq
\label{chi_n}
\chi^{(n)}(0,k)=\sum_{A=1}^4 (i\omega)^n u_A(k)\,.
\eeq
This yields
\beq
u_1=\frac{\omega_2\,\omega_3\,\omega_4\,\chi(0)+i (\omega_2\omega_3+\omega_3\omega_4+\omega_4\omega_2)\dot \chi(0)-(\omega_2+\omega_3+\omega_4)\ddot\chi(0)-i\,  \dddot\chi(0)}{(\omega_2-\omega_1)(\omega_3-\omega_1)(\omega_4-\omega_1)}\,,
\eeq
and similar expressions for the other coefficients $u_A$, up to a permutation of the indices $A$. 

Substituting these coefficients into (\ref{chi_evolution}), one obtains the full time evolution of $\chi(t,k)$, or equivalently $\chi(t,x)$ via inverse Fourier transform. In the generic case where $u_3$ and $u_4$ are nonzero, the imaginary part of $\omega_3$ and $\omega_4$ implies an exponential time evolution, thus signalling an apparent instability. 
As we shall see below, this instability can be avoided by taking into account appropriate boundary conditions, e.g.\ the regularity at spatial infinity.

\subsection{Comparison between the two approaches}
In this subsection, we discuss how the two previous analyses can be reconciled. In the following, we assume that $-1<\alpha<1$ so that the gradient of the background scalar field (\ref{phi_bck}) is time-like. 

First of all, let us note that the background solution (\ref{phi_bck}), given in a non-unitary gauge, can also be described in a unitary gauge by using a new coordinate system, obtained via the boost transformation
\begin{equation}
\label{boost}
\tilde t= \tilde{T}(t,x) \equiv \frac{1}{\sqrt{1-\alpha^2}}(t+\alpha x)\,,\quad
 \tilde x= \tilde{X}(t,x) \equiv \frac{1}{\sqrt{1-\alpha^2}}(x+\alpha t)\,.
\end{equation}
In the coordinates $(\tilde t, \tilde x)$, the scalar field (\ref{phi_bck}) is given by $\bar\phi=\sqrt{1-\alpha^2} \, \tilde t$. 

Accordingly, on substituting
\beq
\omega=\frac{1}{\sqrt{1-\alpha^2}}(\tilde \omega-\alpha\tilde k)\,, \qquad 
k=\frac{1}{\sqrt{1-\alpha^2}}(\tilde k-\alpha\tilde  \omega)\,,
\eeq
which correspond to the (inverse) boost of the wave vector $(\tilde\omega, \tilde k)$, into the dispersion relation (\ref{dispersion}) gives
\beq
\left(1+2\tilde \mu \, \tilde k^2\right)\tilde \omega^2-\tilde k^2=0\,,\qquad \tilde\mu\equiv \mu (1-\alpha^2)^2\,,
\eeq
which is of the form (\ref{dispersion_0}), with a  rescaling of $\mu$ due to the fact that  $\bar\phi$ is not strictly equal   to $\tilde t$, but simply proportional to it. 

\medskip

Similarly, the  equation of motion can be rewritten as
\begin{equation}
\label{eom_unitary}
 \left[ 1 - 2\mu(1-\alpha^2)^2\partial_{\tilde{x}}^2\right]\partial_{\tilde{t}}^2\chi - \partial_{\tilde{x}}^2\chi = 0\,,
\end{equation}
where
\begin{equation}
 \partial_{\tilde{t}} \equiv \frac{1}{\sqrt{1-\alpha^2}}(\partial_t-\alpha\partial_x)\,,\quad
  \partial_{\tilde{x}} \equiv \frac{1}{\sqrt{1-\alpha^2}}(\partial_x-\alpha\partial_t)\,,
\end{equation}
and the perturbation $\chi=\chi(T(\tilde t,\tilde x),X(\tilde t, \tilde x))$ is now viewed as a function of $(\tilde{t},\tilde{x})$ with
\bea
\label{inverse boost}
X(\tilde t,\tilde x) \; = \; \frac{1}{\sqrt{1-\alpha^2}} (\tilde x - \alpha \tilde t) \,, \quad
T(\tilde t, \tilde x) \; = \; \frac{1}{\sqrt{1-\alpha^2}} (\tilde t - \alpha \tilde x) \,. 
\eea
In contrast with the equation of motion written in the original coordinate system, the equation (\ref{eom_unitary}) is only second order in time derivatives.
One can easily decompose this fourth-order equation of motion into two second-order equations, one hyperbolic and  the other elliptic,
\begin{eqnarray}
 (\partial_{\tilde{t}}^2-\partial_{\tilde{x}}^2)\chi & = & \psi\,,\label{eqn:eq-chi}\\
 (\partial_{\tilde{x}}^2-\kappa^2)\psi & = & - \partial_{\tilde{x}}^4\chi\,, \label{eqn:eq-psi}
\end{eqnarray}
where
\begin{equation}
 \kappa = \frac{1}{\sqrt{2\mu}}\frac{1}{1-\alpha^2}\,.
\end{equation}
Hereafter we assume $\mu>0$ so that $\kappa$ is real and positive. 
The second equation (\ref{eqn:eq-psi}) can be easily integrated on a hypersurface where $\tilde t$ is constant (i.e.\ $t+\alpha x$ is constant)
 provided a boundary condition for $\psi$ is properly specified at infinity.  
 Hence, there is clearly no instability in this frame.

\medskip

To understand  how the two approaches are related, let us express one of the complex frequency  modes found above, i.e.   
\beq
\chi_{\rm sh}(t, x)=e^{i (\omega t-  k  x)}\,, \quad {\rm with}\quad \omega=\omega_r\pm i\, \omega_i\,,
\eeq
in terms of the coordinates $(\tilde t, \tilde x)$ gives
\beq
\label{chi_complex}
\chi_{\rm sh}(\tilde t, \tilde x)= \exp\left\{i\left[\frac{(\omega_r+\alpha k)\tilde t-( k+\alpha \omega_r)\tilde x}{\sqrt{1-\alpha^2}}\right]\right\}e^{\mp \frac{\omega_i}{\sqrt{1-\alpha^2}} \tilde t}\, e^{\pm \frac{\alpha\omega_i}{\sqrt{1-\alpha^2}} \tilde x}\,.
\eeq
This shows that these modes diverge  at spatial infinity in the new coordinate system. This explains why these modes do not appear when one starts the analysis around a unitary gauge background and demands the regularity of the initial data at spatial infinity. 

Imposing appropriate boundary conditions for the elliptic equation (\ref{eqn:eq-psi}),  for instance that the acceptable solutions should be well-behaved at spatial infinity,
eliminates the complex frequency modes (\ref{chi_complex}). In the description (\ref{chi_evolution}), such boundary  conditions would impose that  $u_3=u_4=0$.  In terms of initial conditions, this implies that the second and third order derivatives $\chi^{(2)}(0,k)$ and $\chi^{(3)}(0,k)$ are not independent but are instead fixed in terms of    $\chi(0,k)$ and $\dot\chi(0,k)$. Explicitly, one finds
\bea
\label{cond_u_1}
\chi^{(2)}(0,k)&=&\omega_1\, \omega_2\, \chi(0,k)+i(\omega_1+\omega_2)\, \dot\chi(0,k)\,, 
\\
\label{cond_u_2}
\chi^{(3)}(0,k)&=&i \omega_1\, \omega_2(\omega_1+\omega_2)\, \chi(0,k)-(\omega_1^2+\omega_1\,\omega_2+\omega_2^2) \, \dot\chi(0,k)\,.
\eea

One can  check that the single degree of freedom that appears with a  unitary gauge background  automatically verifies the above constraints. Such a mode is of the form
\beq
\chi_u(t,x)=e^{i (\tilde\omega \tilde T(t,x)- \tilde k \tilde X(t,x))}= \exp\left\{i\left[\frac{(\tilde\omega-\alpha\tilde k)t-(\tilde k-\alpha\tilde \omega)x}{\sqrt{1-\alpha^2}}\right]\right\}\,,
\eeq
where $\tilde\omega$ satisfies the unitary gauge dispersion relation (\ref{dispersion_0}). By computing the time derivatives of $\chi_u$, one can verify that the conditions 
(\ref{cond_u_1}) and (\ref{cond_u_2}) are indeed satisfied. 

In summary, we have found that arbitrary initial conditions, defined in some Lorentz frame where the background field is space dependent,  generically lead to the presence of an apparent 
exponential instability. However, this instability is eliminated by imposing appropriate boundary conditions required to solve the elliptic part of the equations of motion.

\subsection{Green's function and emergence of light-cone at long distance}
\label{Green}
In this section, we study further the dynamics of the perturbation $\chi$ in both coordinate systems. 
Let us start considering the equation of motion for  $\chi$ in the $(\tilde t,\tilde x)$ coordinate systems written in the form \eqref{eqn:eq-chi} and \eqref{eqn:eq-psi}. To integrate the first equation  \eqref{eqn:eq-chi} explicitly, 
it is useful to introduce the Green's function $G(\tilde{x},\tilde{x}')$, defined  by 
\begin{equation}
\label{GFeq}
 (\partial_{\tilde{x}}^2 - \kappa^2)G(\tilde{x},\tilde{x}') = \delta(\tilde{x}-\tilde{x}')\,.
\end{equation}
If we require the regularity condition at infinity
\begin{equation}
\lim_{\tilde{x}\to\pm\infty} \left| G(\tilde{x},\tilde{x}')\right|<\infty\,, \label{eqn:bc-G}
\end{equation}
the Green's function \eqref{GFeq} is given by
\begin{equation}
 G(\tilde{x},\tilde{x}') =  -\frac{1}{2\kappa} e^{-\kappa |\tilde{x}-\tilde{x}'|}\,. \label{eqn:Greenfunction}
\end{equation}
As a consequence, the solution to (\ref{eqn:eq-psi}) is 
\bea
\psi(\tilde{t},\tilde{x}) =
 \frac{1}{2\kappa} \int d\tilde{y} \,  \frac{\partial^4 }{\partial\tilde{y}^4}
 \chi \left(T(\tilde{t},\tilde{y}),X(\tilde{t},\tilde{y}) \right) \, 
 e^{-\kappa \vert \tilde{x} - \tilde{y}\vert} \, .
\eea
Substituting this solution back to (\ref{eqn:eq-chi}), one obtains an equation for $\chi$ that includes only second order time derivatives. Obviously, $\psi$ represents the ``shadowy'' mode and the typical length scale of the ``shadow'' is $1/\kappa$. 
In the limit $\kappa \rightarrow \infty$, the length of the ``shadow" vanishes and there is no shadowy mode, as one can see directly
from the Lagrangian \eqref{Lalpha}. 

From the previous analysis, we show that (with appropriate boundary conditions) the equation for $\chi$ in the $(\tilde{t},\tilde{x})$
coordinate system reduces to
\bea
\label{eqforchi}
 (\partial_{\tilde{t}}^2-\partial_{\tilde{x}}^2)\chi \;  = \; \frac{1}{2\kappa} \int d\tilde y \, \frac{\partial^4 }{\partial\tilde{y}^4} \chi \left(T(\tilde{t},\tilde{y}),X(\tilde{t},\tilde{y}) \right) \, 
 e^{-\kappa \vert \tilde{x} - \tilde y\vert} \, .
\eea
Thus, the perturbation $\chi$ is uniquely determined by the datas of $\chi$ and $\partial_{\tilde{t}}{\chi}$ on a constant $\tilde{t}$ hypersurface, say $\tilde{t}=0$.  

\medskip

Now, when one considers the equation of motion in the $(t,x)$ coordinate system, one might wonder whether the values of $\chi$ and $\partial_t {\chi}$ on a constant $t$ hypersurface  can uniquely determine the evolution of the system or not. To answer this question, we first reformulate \eqref{eqforchi} in the $(t,x)$ coordinate system as follows,
\bea
&& (\partial_{{t}}^2-\partial_{{x}}^2)\chi \; =  \; \psi(t,x)  \quad \text{with} \label{eq3}\\
&& \psi(t,x) \; = \; \frac{1}{2\kappa} \int d\tilde y \, \frac{\partial^4}{\partial \tilde y^4}
 \chi \left(T(\tilde{T}(t,x),\tilde y), X(\tilde{T}(t,x),\tilde y) \right) \, 
 e^{-\kappa \vert \tilde{X}(t,x) - \tilde y\vert} \, ,\label{eqn:sol-psi}
\eea
where the functions $T$, $\tilde T$, $X$ and $\tilde X$ 
were given in \eqref{boost} and  \eqref{inverse boost}.
We will argue that the answer is positive, at least for $\kappa L\gg 1$, where $L$ is the length scale of interest
which characterizes the variations of $\chi$ in space.
For $\kappa L\gg 1$, we also argue that the concept of lightcone emerges. 

The absolute value of the Green's function (\ref{eqn:Greenfunction}) has the maximum $(2\kappa)^{-1}$ at $\tilde x=\tilde x'$ and decays exponentially away from it.
Hence, if the length scale $L$ 
is sufficiently longer than $1/\kappa$ then, as one can easily confirm for each Fourier mode, (\ref{eqn:sol-psi}) implies that 
$\psi$ scales as 
\begin{equation}
 |\psi| \sim \mathcal{O}(\epsilon)\times |\partial_X^2\chi| \ll |\partial_X^2\chi|\,,
\end{equation}
where we have introduced the small bookkeeping parameter $\epsilon=1/(\kappa L)$.
Therefore, at the lowest order in $\epsilon$, (\ref{eqn:eq-chi}) reduces to
\begin{equation}
\label{eqn:eq-chi0}
 (\partial_t^2-\partial_x^2)\chi \simeq 0\,,
\end{equation}
which gives an approximate solution $\chi\simeq \chi^{(0)}(t,x)$ from the initial values of $\chi$ and $\partial_t {\chi}$ on a hypersurface of 
constant $t$. For this approximate solution the concept of lightcone makes sense 
(as we recover the usual d'Alembert equation).

Furthermore, one can systematically improve the approximation by expanding $\chi$ and $\psi$ in powers of 
$\epsilon$ as
\begin{equation}
 \chi = \chi^{(0)}+\chi^{(1)}+\chi^{(2)}+\cdots\,,\quad
  \psi = \psi^{(1)}+\psi^{(2)}+\cdots\,, \label{eqn:long-distance-expansion}
\end{equation}
where $\chi^{(n)}=\mathcal{O}(\epsilon^n)$ and $\psi^{(n)}=\mathcal{O}(\epsilon^n)$. Substituting these expansions in (\ref{eq3}) and (\ref{eqn:sol-psi}), at the lowest order in $\epsilon$, one recovers \eqref{eqn:eq-chi0} for $\chi^{(0)}$
and
\begin{equation}
 \psi^{(1)}(t,x) = \frac{1}{2\kappa}\int d\tilde{y} \, 
  \frac{\partial^4}{{\partial \tilde{y}}^4}\chi^{(0)}\left(T(\tilde{T}(t,x),\tilde{y}),X(\tilde{T}(t,x),\tilde{y})\right) \, 
 e^{-\kappa \vert \tilde{X}(t,x) - \tilde y\vert} \,. \label{eqn:sol-psi0}
\end{equation}
Suppose that the initial condition for $\chi$ is specified on an initial surface at $t_0$ as ($\chi(t_0,x)$, $\partial_t {\chi}(t_0,x)$) $=$ ($\chi_0(x)$, $\chi_1(x)$). One can easily solve (\ref{eqn:eq-chi0}) for $\chi^{(0)}$ with the initial condition given  by 
($\chi^{(0)}(t_0,x)$, $\partial_t{\chi}^{(0)}(t_0,x)$) $=$ ($\chi_0(x)$, $\chi_1(x)$), and obtain a solution $\chi^{(0)}(t,x)$ for all $(t,x)$. One can then calculate the right hand side of (\ref{eqn:sol-psi0}) 
to give $\psi^{(1)}(t,x)$ for all $(t,x)$. The leading correction to $\chi^{(0)}$ is given by solving the $\mathcal{O}(\epsilon)$ part of (\ref{eq3}), namely
\begin{equation}
 (\partial_t^2-\partial_x^2)\chi^{(1)} = \psi^{(1)}\,,
\end{equation}
with the initial condition $\chi^{(1)}(t_0,x)=\partial_t{\chi}^{(1)}(t_0,x)=0$. Higher order corrections are also calculable in a similar way. The derivative (or long-distance) expansion (\ref{eqn:long-distance-expansion}) is expected to converge as far as $\epsilon\ll 1$.

We thus conclude that for $\mu>0$ and under the appropriate boundary condition, the values of $\chi$ and $\partial_t{\chi}$ on a 
surface of constant $t$  (instead of $\tilde{T}(t,x)$ constant) uniquely determines the evolution of the system as far as the length scale of interest is sufficiently longer than $1/\kappa$. Moreover, in this limit, since $\chi^{(0)}$ gives a good approximation to the full solution $\chi$ and the concept of lightcone makes sense for $\chi^{(0)}$, we also conclude that the concept of lightcone emerges at long distances. In summary, if we are interested in physics at length scales sufficiently longer than the length of the ``shadow'' then the ``shadowy'' mode is invisible and the evolution of the system appears to be Lorentz-invariant.

\section{Conclusion}
We have studied Higher-Order Scalar-Tensor theories that are not DHOST theories but are nevertheless degenerate when restricted to the unitary gauge. These theories, which we have dubbed U-degenerate,  appear to contain one more dynamical degree of freedom in their covariant formulation than when restricted to the unitary gauge. 

In the first part of the present work, we have shown  how the class of theories that are either DHOST or U-degenerate can be systematically classified. We  have found that quadratic  theories of this class can be described by a Lagrangian that depends on five arbitrary functions (see Eq (\ref{L_Ud_quad})), obtained by combining two DHOST Lagrangians: the first includes the curvature term (and is totally U-degenerate), the second can be written in a simple form where the degeneracy is manifest.  We have then extended this description  to theories that are cubic and higher order. Note that all our general Lagrangians that describe this class of theories also include as particular cases DHOST theories, since the latter automatically satisfy the unitary-gauge degeneracy condition, as a consequence of the full system of degeneracy conditions.

In the second part of this article, we have tried to reconcile  the apparently contradictory points of view when the background scalar field 
(whose gradient is assumed to be time-like) is described in  the unitary gauge or in a different  gauge, by studying  a simple toy model  where the tensor modes, i.e. gravity, are ignored. In this model, we have found that the extra degree of freedom that appears in a non-unitary gauge can be understood as a generalized instantaneous  mode, or ``shadowy mode", which does not propagate. Indeed, this extra mode is governed by an elliptic equation, which is manifest in the unitary gauge (although somewhat obscured by the mixing of time and space in another gauge). Imposing appropriate boundary conditions, namely regularity at spatial infinity,  leads to the elimination of the apparent instability in  a non-unitary gauge. In this sense, the fact that the system in the unitary gauge seems to contain one less dynamical degree of freedom than in another gauge is due to the fact that the boundary conditions are already implemented implicitly in the unitary gauge, whereas they need to be taken into account explicitly in the other gauges.

Beyond the particular example we have studied, our analysis strongly suggests that U-degenerate theories, 
when the scalar field gradient is time-like and with appropriate boundary conditions,
 propagate a single scalar degree of freedom, while the extra degree of freedom, the shadowy mode, is non-dynamical. 

 This would mean that, within these conditions, U-degenerate theories are safe from Ostrogradski instabilities and  
 therefore worth exploring phenomenologically. The behaviour of U-degenerate theories  might  differ from that of DHOST theories\footnote{For linear cosmological perturbations, one can find in  \cite{Langlois:2017mxy}  the quadratic action for the physical scalar degree of freedom  when taking into account only the unitary degeneracy condition.},  and we plan to investigate their potentially new features in the future. It would also be very interesting to extend our analysis to the case where the tensor degrees of freedom are taken into account, studying for example the linear perturbations about a non-isotropic cosmological background.  

Another important issue related to the presence of the shadowy/instantaneous mode is the existence of black holes. In  khronometric theories, it was shown that black holes still exist, but their boundaries are  now  
universal horizons~\cite{Blas:2011ni}. Recently, such black holes have been studied intensively \cite{Berglund:2012fk,Wang:2017brl}, and it would be very interesting to investigate the existence, formation  and thermodynamics of black holes in   U-degenerate theories.

\section*{Acknowledgements}
KN and AW wish to thank warmfully the YITP in Kyoto (where this article was initiated and later completed) for its hospitality. KN wants also to thank Y. Masahide for very interesting discussions on this topic during its stay at Tokyo Institute of Technology. 
DL would like to thank the Yukawa Institute for Theoretical Physics at Kyoto University, where this work was completed during the YITP symposium YKIS2018a  ``General Relativity -- The Next Generation --'' and  the workshop YITP-T-17-02 "Gravity and Cosmology 2018".
We also thank  Gilles Esposito-Farese for very  instructive discussions on this topic.
ADF was supported by JSPS KAKENHI Grant Numbers 16K05348, 16H01099. The work of SM was supported by Japan Society for the Promotion of Science (JSPS) Grants-in-Aid for Scientific Research (KAKENHI) No. 17H02890, No. 17H06359, No. 17H06357, and by World Premier International Research Center Initiative (WPI), MEXT, Japan. The work of AW is supported in part by  the National Natural Science Foundation of China (NNSFC), Grant Nos. 11375153 and  11675145.
\appendix 

\section{Instantaneous modes in khronometric theories}
Instantaneous modes have been introduced in the context of massive gravity in \cite{Gabadadze:2004iv},  and later on considered 
in \cite{Blas:2010hb,Blas:2011ni} in the context of khronometric theories which are simple examples of higher-order scalar-tensor theories. 
In the later case, the notion of instantaneous modes has been defined  at the perturbative level about a homogeneous background. 

Indeed, if one considers the dynamics of a small perturbation of the scalar field (the khronon), $\phi=t+\chi$ where
$\chi$ is the perturbation about the solution $\phi=t$, on a fixed Minkowski background $g_{\mu\nu}=\eta_{\mu\nu}$, 
one easily sees that the quadratic action for $\chi$ is higher order in space derivatives only:
\bea
S_{khr}^{(2)}[\chi]\; = \; \int d^4x \, \left[ \alpha (\partial_i \dot \chi )^2 + \beta (\Delta \chi)^2\right] \, ,
\eea 
where $\alpha$ and $\beta$ can be reduced to non-zero constants (not functions) for our discussion here.
The corresponding  dispersion 
relation 
\bea
\alpha \omega^2 k^2 \, + \, \beta k^4 \; = \;  \alpha k^2(\omega^2 -c_s^2 k^2 )\; = 0 \,,
\eea
shows that the action describes a single mode propagating with a finite velocity $c_s^2=-\beta/\alpha$. However, one can interpret the presence of 
higher spatial derivatives  in the equations of motion as the signature of a second mode propagating with an infinite speed: this is the reason why 
such a mode has been said to be instantaneous in  \cite{Blas:2010hb}. 
Obviously, these instantaneous modes are simple examples of our shadowy modes.
 
\section{A covariant form for the totally degenerate action}

This section aims at formulating the Lagrangian \eqref{LU0} in a covariant form. For that purpose,
we start using the (scalar) Gauss-Codazzi relation 
\bea
\label{GC}
R \; = \; {}^3\! R[h] + K_{\mu\nu} K^{\mu\nu} - K^2 - 2 \nabla_\mu (\acc^\mu - K n^\mu) \, ,
\eea
which links the four dimensional Ricci scalar $R$ (associated to the metric $g_{\mu\nu}$)
with the three dimensional Ricci scalar ${}^3\!R[h]$ associated to the three-dimensional spatial
metric
\bea
h_{\mu\nu} \equiv g_{\mu\nu}- \frac{1}{X} \, \phi_\mu \phi_\nu \, .
\eea
In \eqref{GC}, the normal unit vector is given by 
\bea
n_\mu \; \equiv \; \frac{\phi_\mu}{\sqrt{-X}} \, ,
\eea
from which we easily deduce the components of the acceleration vector $\acc_\mu$ and of the
fundamental two-form $K_{\mu\nu}$
\bea
\acc_\mu & = & n^\nu \nabla_\nu n_\mu =  -\frac{1}{X} h_{\mu\alpha} \phi^{\alpha \nu} \phi_\nu  = 
- \frac{1}{X} \left[ \phi^\nu \phi_{\mu\nu} - \frac{1}{X} (\phi^\alpha \phi_{\alpha\beta} \phi^\beta) \phi_\mu\right] \, , \\
K_{\mu\nu} & = & h_\mu^\alpha h_\nu^\beta \nabla_\alpha n_\beta = 
\frac{1}{\sqrt{-X}} \left[ \phi_{\mu\nu} + \frac{1}{X^2} (\phi^\alpha \phi_{\alpha\beta} \phi^\beta) \phi_\mu\phi_\nu
-  \frac{1}{X} \phi^\beta (\phi_\mu \phi_{\nu\beta} + \phi_\nu \phi_{\mu\beta})\right]\, .
\eea
Replacing these expressions in the (scalar) Gauss-Codazzi relation \eqref{GC}, one easily shows, after an immediate
calculation, that $L_\td[f_2, \aq_5] $ can be expressed as
\bea
\label{L0simp}
L_\td[f_2, \aq_5]  \; = \; f_2 \, {}^3\!R[h] + \alpha \, \acc^2 + 2 f_{2\phi} \, h^{\mu\nu} \phi_{\mu\nu} +\nabla_{\mu}\left[\frac{2f_2}{X}\left(\phi_{\nu}\phi^{\mu\nu}-\phi^{\mu}\Box\phi\right)\right]\, ,
\eea
where $\acc^2\equiv \acc_\mu \acc^\mu$, and $\alpha\equiv -X^3 \aq_5$ is a function independent of $f_2$. 

\medskip

The expression \eqref{L0simp}
is interesting because it shows explicitly (and in a covariant way) that $L_\td$ does not contain any second time derivatives of the scalar
field. Second derivatives are space-like only. First of all, the original four dimensional Ricci scalar combines with the Lagrangians
$L_A^{(2)}$ in order to reduce to the three-dimensional Ricci scalar of $h_{\mu\nu}$, which is the first term in  \eqref{L0simp}, plus a small number of additional terms at the end of the calculation. Then, the second term in \eqref{L0simp}, constructed from the acceleration vector, can be easily reformulated as follows:
\bea
\label{second term}
\alpha \, \acc^2 \; = \; \frac{\alpha}{4X^2} (\partial_\alpha X) h^{\alpha\beta} (\partial_\beta X) \; = \; 
 (\partial_\alpha F) h^{\alpha\beta} (\partial_\beta F) \, ,
\eea
where $F(\phi,X)$ is a function of $\phi$ and $X$ such that $F_X^2=\alpha/(4X^2)$.
Even though $F$ contains time derivatives of $\phi$ via $X$, \eqref{second term} does not produce higher time derivatives, because
only space derivatives of $F$ are present. Finally, the third term in \eqref{L0simp} involves also  first time derivatives
of $\phi$ only which appear via the Christoffel symbol $\Gamma_{\mu\nu}^\rho$ of the metric $g_{\mu\nu}$ according to
\bea
h^{\mu\nu}\phi_{\mu\nu} \,  \ni \, -h^{\mu\nu}\Gamma^{0}_{\mu\nu}\dot{\phi} \, .
\eea
Of course, we disregard the last term in \eqref{L0simp} which is an irrelevant total derivative.

\section{Counting degrees of freedom}
\label{dofinst} 
In this section, we count the number of degrees of freedom of the totally degenerate theory \eqref{LU0} expressed in the unitary gauge
thanks to a Hamiltonian analysis. To do so, we first rewrite the action  \eqref{LU0}  as follows
\bea
\label{SOU1}
 \int d^4x \sqrt{\gamma} \left[ f(N) \, {}^3\!R + \beta(N) \gamma^{ij}\dot\gamma_{ij} + \tilde{N}^i \partial_i N\right] \, ,
\eea
where
\bea
f(N) \, \equiv \, N f_2(N) \, , \quad
\beta(N) \, \equiv \, - \frac{f_{2\phi}(N)}{N} \, , \quad
\tilde{N}^i \, \equiv \, 2\frac{\partial \beta}{\partial N} N^i \, + \, \frac{\alpha_5}{N^7} \gamma^{ij} \partial_j N \, .
\eea 
For simplicity we have omitted to mention explicitly the time dependence of the functions in the Lagrangian. 
When $(\partial \beta/\partial N) \neq 0$ (what we assume here), one can change the variable $N^i$ by $\tilde{N}^i$. 
Integrating out this new  variable, one obtains that 
$N$ is a function of time only, and then the action \eqref{SOU1} is shown to be equivalent to
\bea
 \int d^4x \sqrt{\gamma} \left[ f(t) \, {}^3\!R + \beta(t) \gamma^{ij}\dot\gamma_{ij} \right] \, ,
\eea
where we have used the notation $f(t)$ for $f(t,N(t))$ (same thing for $\beta(t)$). 

To start the Hamiltonian analysis, one introduces  the 6 pairs of conjugate variables ($\gamma_{ij}$, $p^{kl}$) with the Poisson bracket
\bea
\{\gamma_{ij}(\vec{x}),p^{kl}(\vec{y}) \} = \delta_{(i}^k \delta_{j)}^l\delta^3(\vec{x}-\vec{y}) \, ,
\eea
which satisfy the 6 primary constraints
\bea
\chi^{ij} \equiv p^{ij} -\beta(t)\sqrt{\gamma}\gamma^{ij} \approx 0 \, .
\eea
To go further, it is very useful to decompose the family of primary constraints into two independent sets
$(\chi^{ij}) = (\chi^i_\parallel,\chi^i_\perp)$ where $\chi^i_\parallel$ are the 3 longitudinal components of the constraints
\bea
\chi^i_\parallel \; \equiv \; D_j\left(\frac{\chi^{ij}}{\sqrt{\gamma}}\right)\, = \, D_j \left(\frac{p^{ij}}{\sqrt{\gamma}}\right) \, ,
\eea
and $(\chi^i_\perp)$ are the 3 transverse components. Thus, the total Hamiltonian reads
\bea
H_{tot} = \int d^3 x \, \sqrt{\gamma} \left[ -f(t) \, {}^3\! R + \lambda_i \chi^i_\perp  + \mu_i \chi^i_\parallel  \right] \, ,
\eea
where $\lambda_i$ and $\mu_i$ are Lagrange multipliers which enforce the primary constraints. 
It is easy to see that $\chi^i_\parallel \approx 0$ are always conserved under time evolution whereas the conservation
of $\chi^i_\perp  \approx 0$ leads to 3 secondary constraints $\varphi_i \approx 0$. To see this in indeed the case, let us remark that
\bea
\dot \chi_{ij} \; = \; \frac{\partial \chi_{ij}}{\partial t} + \{\chi_{ij},H_{tot}\} \; = \; f(t)G^{ij} - \dot\beta(t) \gamma^{ij} \; \approx \; 0 \, ,
\eea
where $G^{ij}$ are the component of the Einstein tensor associated to $\gamma_{ij}$. Now, it becomes obvious (due to the conservation
of $G^{ij}$) that $\dot \chi^i_\parallel \approx 0$ with no conditions, and only three components of $\dot \chi_{ij} $ are non-vanishing,
which leads to 3 secondary constraints. 

The Dirac algorithm closes here (there is no
tertiary constraints) with 9 constraints in total: $\chi^i_\parallel \approx 0$ are in fact first class (and they are associated to the invariance of the theory under spatial diffeomorphisms); the 6 remaining constraints form a set of second class constraints. 
As we started with 6 pairs of variables, we end up with $[6 - 3 - 6/2]=0$ degree of freedom. 

\medskip

In the special case where $(\partial\beta/\partial N)=0$, which means that $\beta$ depends on $t$ only, the action \eqref{SOU1} reduces
to
\bea
S \; = \; \int d^4x \sqrt{\gamma} \left[ f(N) \, {}^3\!R + \beta(t) \gamma^{ij}\dot\gamma_{ij} + \tilde{\alpha}(N) \gamma^{ij}\partial_i N
\partial_j N \right] \, ,
\eea
where $\tilde\alpha(N)=\alpha_5(N)/N^7$. To make the Hamiltonian analysis, we start now with 7 pairs of conjugate variables
\bea
\{\gamma_{ij}(\vec{x}),p^{kl}(\vec{y}) \} = \delta_{(i}^k \delta_{j)}^l\delta^3(\vec{x}-\vec{y}) \, , \quad
\{N(x),\pi(y)\} = \delta^3(\vec{x}-\vec{y}) \, ,
\eea
which satisfy the 7 primary constraints
\bea
\chi^{ij} \equiv p^{ij} -\beta(t) \gamma^{ij} \approx 0 \, , \quad \pi \approx 0 \, .
\eea
The analysis of the constraints $\chi^{ij}$ is exactly the same as  the previous case. Concerning the new constraint $\pi \approx 0$,
its time evolution leads to the secondary constraint 
\bea
{\cal H} \; \equiv \; \frac{\delta S}{\delta N} \; \approx \; 0 \, ,
\eea
which is nothing but the Euler-Lagrange equation for the lapse $N$. There are no tertiary constraints and we end up with 3 first class
constraints (associated to the invariance under space diffeomorphisms) together with 8 second class constraints. As we have started with 7 
pairs of conjugate variables, here again we conclude that the theory has no degrees of freedom.

\bibliographystyle{utphys}
\bibliography{Ugauge}

\end{document}